# Scaling Pedestrian Crossing Analysis to 100 U.S. Cities via AI-based Segmentation of Satellite Imagery

Marcel Moran[a,*], Arunav Gupta[b], Jiali Qian[b], Debra Laefer[a,c]

[a]Center for Urban Science & Progress, NYU Tandon School of Engineering.
370 Jay St, 13th Floor, Brooklyn, NY, USA, 11201. marcel.moran@nyu.edu

[b]Center for Data Science, NYU Graduate School of Arts & Sciences
60 5th Ave, New York, NY, 10011. arunav.gupta@nyu.edu, tony.qian@nyu.edu

[c]Department of Civil and Urban Engineering, NYU Tandon School of Engineering
6 MetroTech Center, Rogers Hall RH 400, Brooklyn, NY 11201. debra.laefer@nyu.edu

Accurately measuring street dimensions is essential to evaluating how their design influences both travel behavior and safety. However, gathering street-level information at city-scale with precision is difficult given the quantity and complexity of urban intersections. To address this challenge in the context of pedestrian crossings – a crucial component of walkability – we introduce a scalable and accurate method for automatically measuring crossing distance at both marked and unmarked crosswalks, applied to America's 100 largest cities. First, OpenStreetMap coordinates were used to retrieve satellite imagery of intersections throughout each city – totaling roughly three million images. Next, Meta's Segment Anything Model was trained on a manually-labelled subset of these images to differentiate drivable from non-drivable surfaces (i.e. roads vs. sidewalks). Third, all available crossing edges from OpenStreetMap were extracted. Finally, crossing edges were overlaid on the segmented intersection images, and a grow-cut algorithm was applied to connect each edge to its adjacent non-drivable surface (e.g. sidewalk, private property, etc.), thus enabling the calculation of crossing distance. This achieved 93% accuracy in measuring crossing distance, with a median absolute error of 2 feet 3 inches (0.69 meters), when compared to manually-verified data for an entire city. Across the 100 largest U.S. cities, median crossing distance ranges from 32 feet to 78 feet (9.8 – 23.8m), with detectable regional patterns. Median crossing distance also displays a positive relationship with cities' year of incorporation, illustrating in a novel way how American cities increasingly emphasize wider (and more car-centric) streets.

# 1. Introduction

Making informed planning, design, and construction decisions regarding streets requires accurate information that is both granular at the block level and comprehensive to the municipal scale. Presently, datasets on intersection layouts are often missing, outdated, incomplete, and/or not machine readable. A fundamental component of intersections are pedestrian crossing distances, which influence both pedestrian safety and walkability (Martin 2006; Abley et al. 2011). Thus, being able to generate accurate, citywide datasets of crossing distances can empower urban planners and traffic engineers to visualize patterns related to this feature of the built environment and strategically target streetscape investments. Optimizing the location of pedestrian refuge islands and sidewalk extensions are particularly important for ensuring that urban streets become more walkable, especially for persons with mobility impairments, parents with strollers, and older adults, who may be most at risk within and deterred by long crossings (Wilmut and Purcell 2022). That the U.S. is in the midst of a significant increase in pedestrian fatalities (following years of declines), makes such work particularly relevant (Schneider 2020; Cova 2024).

Scholarship in the last two decades has established an association between longer crossing distances and increased crashes. For example, an analysis of 10 years of pedestrian crashes at roughly 1,600 intersections in Utah found that for every 12 additional feet (3.7m) of crossing distance, pedestrian crashes increased by 5% (Islam et al. 2022). In Oregon, review of crash data between 2007-2014 at 191 crossings revealed that as the number of traffic lanes of the street increased, so did the risk-ratio for pedestrian crashes (Monsere et al. 2016). Crossing distances also impact more than just safety; in evaluating what aspects of streets in Mumbai, India were relevant to pedestrian flow, Kadali and Vedagiri (2015) determined that increasing crossing distance decreased intersections' perceived walkability. In the field of transportation planning, leading street-design guides advocate for crossing distance to be as short as possible, to make walking both safer and more likely (Duncan et al. 2016; Feuer et al. 2020).

For such reasons, urban scholars increasingly pursue means to identify and measure pedestrian crossings at larger scales. Recently, Verma and Ukkusuri (2023) mapped urban crosswalks with high accuracy in four American cities by training a deep-learning model to identify crosswalk markings from satellite imagery. That study extended an earlier computer-vision based attempt at identifying crosswalks by Ahmetovic and colleagues (2017) and complements the work of Li and colleagues (2023), who similarly trained a computer-vision model to recognize crosswalks from street-view imagery. Critically, those studies primarily rely on crossing *markings* (e.g. white or yellow 'zebra' markings) as a primary means for identification, which raises the importance of methodologies that can also capture *unmarked* crossings, which can comprise a large percentage of total crossings in American cities, such as at 42% of intersections in San Francisco (Moran 2022). In this vein, Moran and Laefer (2024) used a combination of data programmatically extracted from OpenStreetMap, along with manually-labeled satellite imagery to measure crossings (both marked and unmarked) across three cities, and found both intra- and inter-urban patterns that likely discourage walking in spatially-distinct ways. Though highly accurate, that approach was labor intensive and time consuming, thus hampering its scalability.

More broadly, employing satellite imagery to study urban forms has grown significantly, as image resolution and availability have increased. In the realm of surface transportation, such efforts have included identification of roads, on-street parking, and traffic patterns (Haverkamp 2002; Larsen et al. 2009; Moran 2020). Computer-vision models' increasing ability to segment images has led to their increased use by urban researchers (Albert et al. 2017; Bagwari et al. 2023), including Meta's Segment Anything Model (SAM) (Ren et al. 2024; Osco et al. 2023). Building on these converging strands of research, the goal of this study is to leverage both street-network information embedded within OpenStreetMap, in combination with Meta's SAM, to automate accurate measurement of marked and unmarked pedestrian crossing distances at city-scale, thereby unlocking new insights on pedestrian networks nationally.

**2. Methodology**

The scope of this study is to automatically measure pedestrian crossing distances throughout the 100 largest U.S. cities by population (see *Supplementary Materials S1* for list of cities). This requires maintaining accuracy across municipalities that vary in size, urban form (e.g. dense and walkable vs. sprawling and car-centric), topography, building height (and resulting shade), and tree cover. In addition, to ensure this method's reproducibility and potential for its adaptation to other urban questions, only freely-available imagery and models are employed, and reasonable processing time (i.e. roughly one hour per city) was prioritized. Guided by these goals, the analysis is comprised of three primary steps (see **Figure 1**):

1. **Intersection Image Segmentation:** Train Meta's Segment Anything Model (SAM) to differentiate between drivable and non-drivable surfaces on satellite-imagery tiles, centered on street intersections.
2. **Crossing Edge Extraction**: Extract all available pedestrian crossing edges from OpenStreetMap, and generate additional edges by expanding crossing nodes into edges.
3. **Grow-Cut Algorithm**: Overlay crosswalk edges extracted from OpenStreetMap on segmented satellite imagery and deploy a grow-cut algorithm on each edge, resulting in accurately-measured pedestrian crossings.

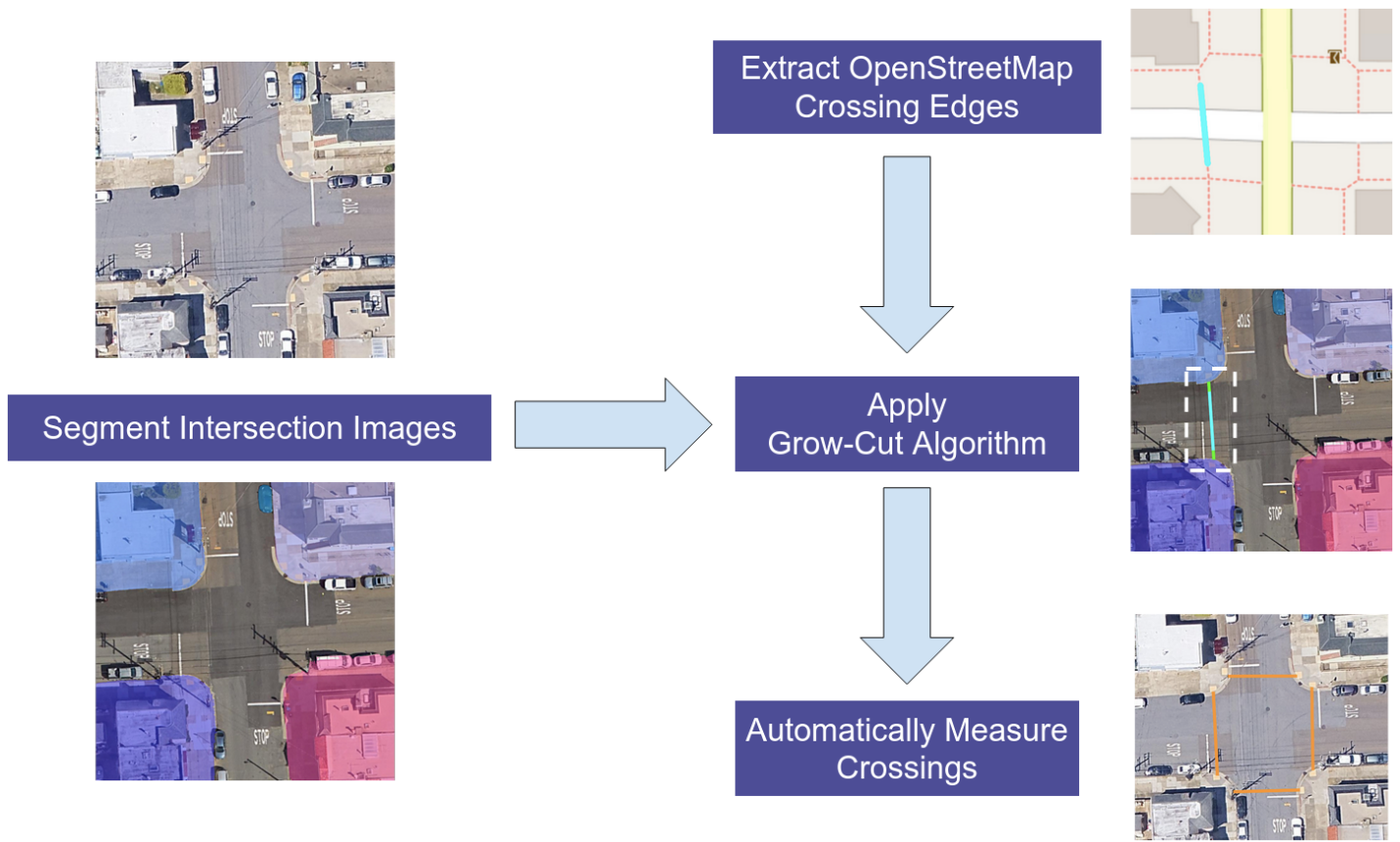

**Figure 1:** Diagram of analysis steps, including (1) segmentation of intersection satellite imagery, (2) extraction of crossing edges from OpenStreetMap, and (3) application of the grow-cut algorithm to the composite data set.

**2.1 Intersection Image Segmentation**

First, the Python package OSMnx (Boeing 2017) was used to generate the latitude-longitude coordinates of all street intersections and mid-block crossings for the 100 most populous cities in the United States. Those coordinates were then used to retrieve satellite-imagery tiles from Google Maps API, centered on each intersection. Each resulting image encompasses a 25-meter radius from its center point and is 1629 x 1629 pixels, with each pixel representing a 30x30cm patch on the ground. These images capture a wide range of intersection sizes and layouts, and include contextual information from surrounding land uses (e.g. street trees, buildings, sidewalks, etc.). In total, approximately three million satellite imagery tiles were queried, totaling 8.3 terabytes of data. Tile retrieval was implemented with a parallelized approach that utilized multiple "serverless" workers simultaneously fetching tiles from different parts of a city, enabling a peak retrieval rate of 2,000 tiles per minute. Chicago required the largest number of tiles to be fetched, roughly 80,000, which took approximately 40 minutes.

After compiling these intersection tiles, the core computer-vision task was segmenting non-drivable surfaces (e.g. sidewalks, buildings, parks, and pedestrian-refuge islands). This was critical because accurately measuring crossing distance entails determining a crossing's extent from one non-drivable segment to another (i.e. from sidewalk to sidewalk or from sidewalk to refuge island). As a 'zero-shot' model, Meta's SAM 2.1 (hereafter, SAM; Ravi et al. 2024), can segment objects without training data, but its performance can be improved via fine-tuning with manually-labeled samples (Ma et al. 2024). To begin, satellite images of street intersections from 14 cities – covering a wide range of densities and urban form – were manually annotated to include masks for all of the non-drivable surfaces, while leaving drivable surfaces (e.g. roads) unlabeled (see **Figure 2**).

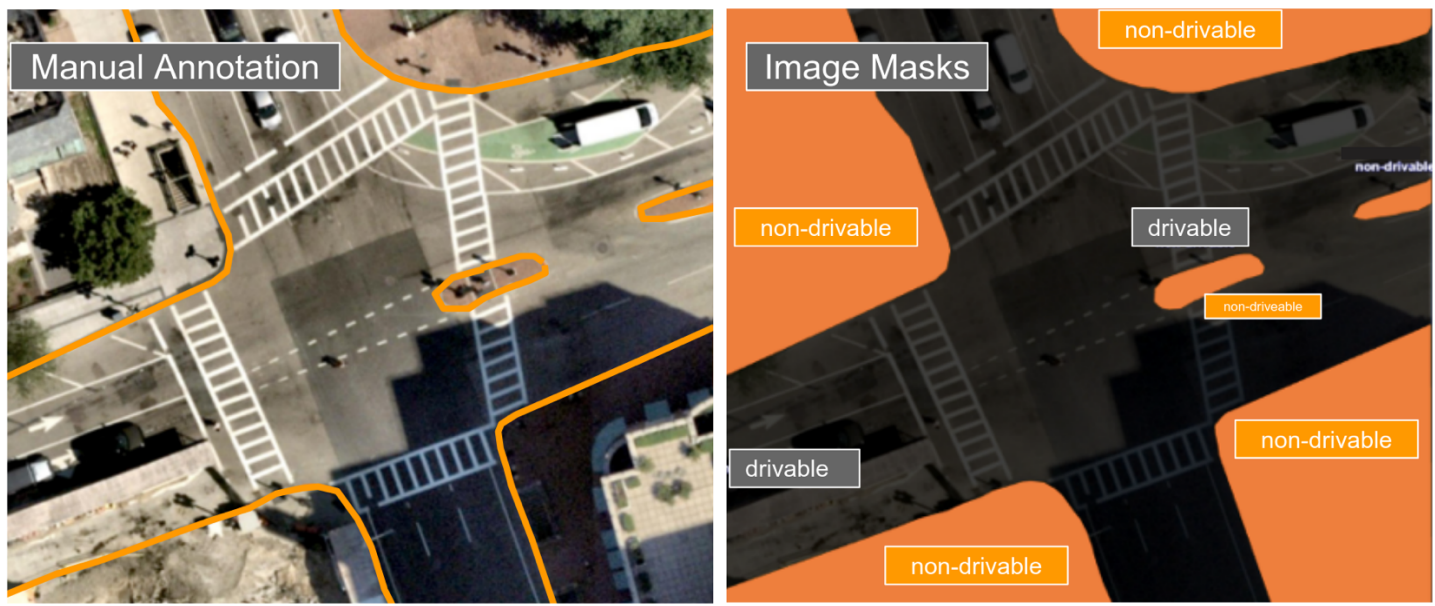

**Figure 2:** Manually-annotated intersection image (left), with masks for non-drivable surfaces and the same intersection viewed as mask polygons (right).

Instead of manually labeling images at random, high-entropy intersection images were prioritized, as these contained greater visual complexity (see **Figure 3**). In contrast, low-entropy images typically depicted more homogeneous locations such as parking lots and industrial areas, which contain only limited training value for distinguishing non-drivable surfaces. Entropy was calculated using the standard log-probability sum (Shannon 1948), where the probability distribution was defined using grayscale image pixel values rescaled to a 0-1 range. Tests using all three-color channels (red, green, blue) to define the pixel-value distribution resulted in no significant difference in the entropy result, so gray-scaling was used for computational efficiency. When plotting all intersection images for a given city by entropy, a long left-tail is present, which represents images less suitable for training.

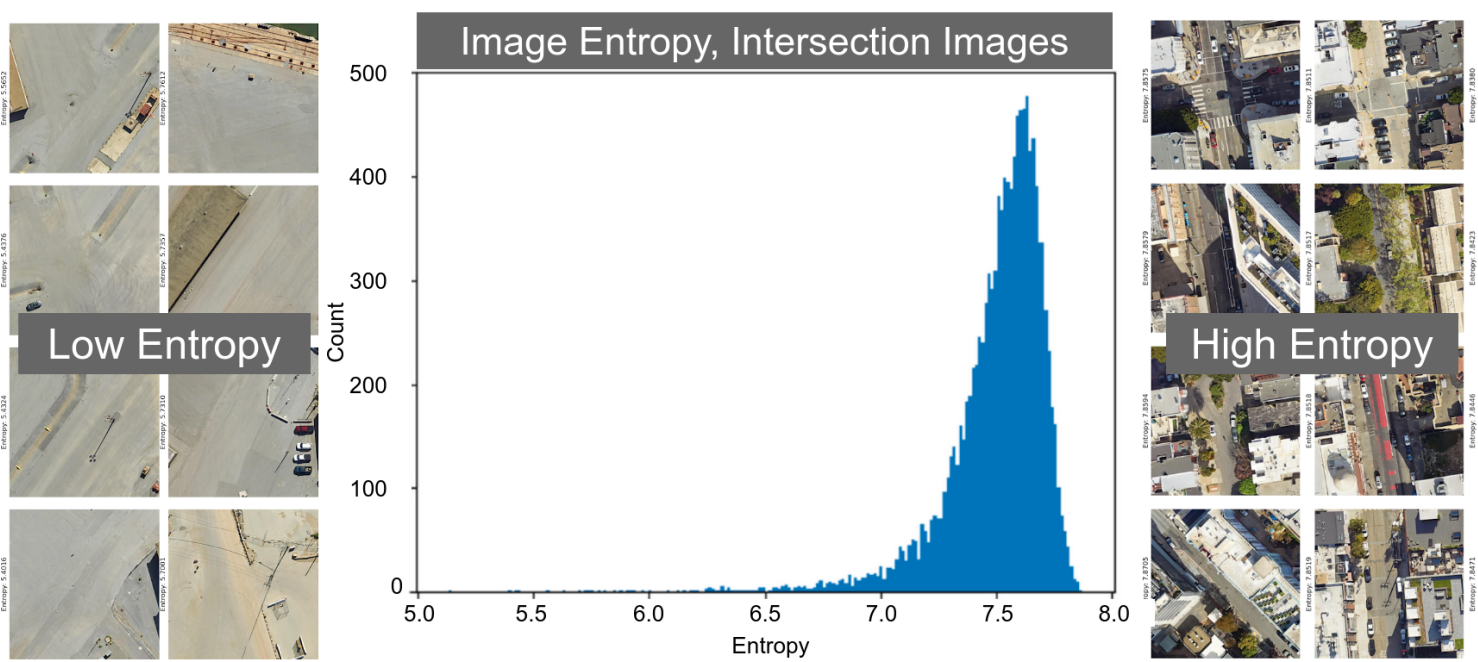

**Figure 3:** Entropy distribution for satellite imagery of street intersections in San Francisco (center), with examples of low-entropy images (left), and high-entropy images (right).

For fine-tuning SAM, 193 total high-entropy intersection images were selected for manual annotation. To increase the size of this training set, all manually-annotated images were horizontally and vertically flipped, rotated (clockwise, counter-clockwise, and upside down), Gaussian blurred (up to 1.5px), and modified in terms of random noise (up to 1.05% of pixels). These modifications artificially expanded the training set to a final collection of 5,790 images.

Following this fine-tuning, SAM produced pixel-level masks of non-drivable surfaces within each of the three million intersection images drawn from all 100 cities (Dwyer et al. 2024).

To enable this, each intersection image was first resized to 1024x1024 pixels to meet the model's input requirements. Next, each image was converted into a 16x16 point grid, wherein each point serves as an 'anchor' for SAM to view the image. SAM starts at each of these anchor points and gradually expands outward to form potential boundaries for the objects for which it is trained to segment. During fine-tuning, a denser point grid (24x24) in SAM was tested but produced more erroneous segmentation of objects in the roadway (i.e. false positives), for objects such as parked cars. As the point-grid density heavily influences the computational load of image segmentation, a 16x16 point grid was found to produce the best balance between accuracy and processing speed.

As part of the segmentation process, masks of non-drivable spaces were post-processed to extract contours that were then converted to polygonal geometries (see **Figure 4**). To maintain spatial accuracy, each segmented image's file name was encoded with its center-point coordinates, which was later used to compute the corresponding bounding box. This allowed the resulting polygons to be mapped back to their real-world geographic locations, thereby ensuring alignment with original satellite imagery.

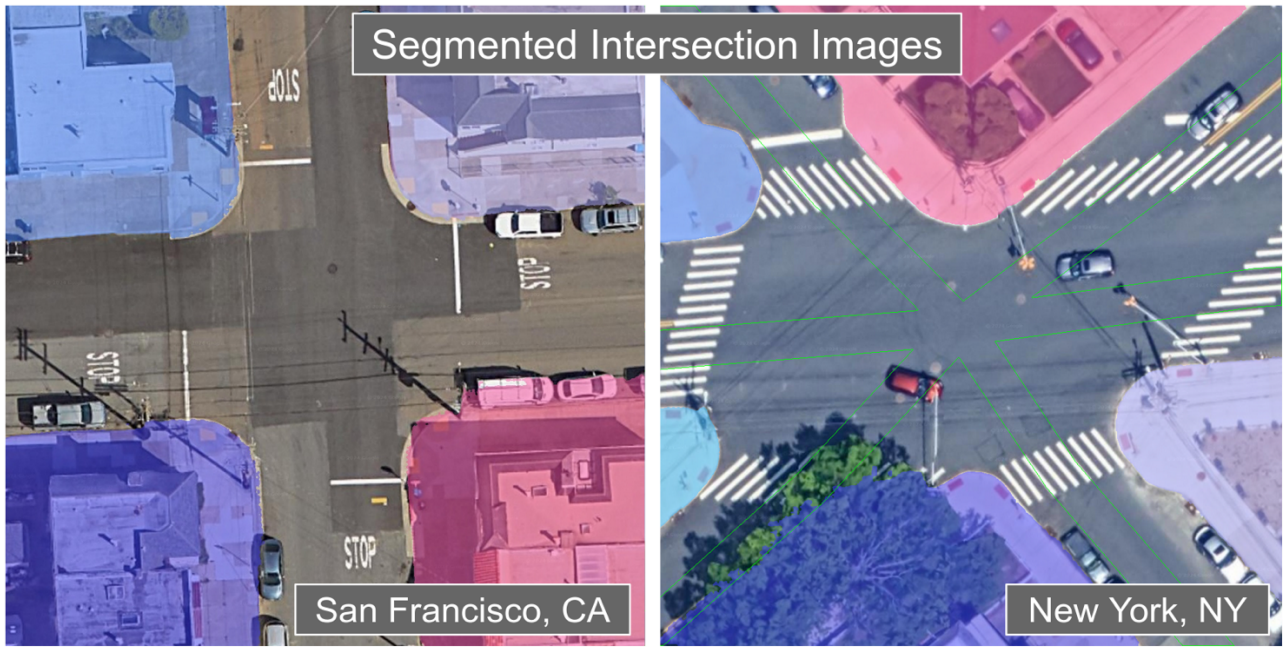

**Figure 4**: Segmentation of intersection images for non-drivable surfaces by SAM.

Using the geographic bounding boxes derived from filenames, the mask contours were mapped to latitude and longitude coordinates and transformed into polygons. Next, for each image, the inverse selection of these polygons was used to identify all *drivable* areas. These inverted polygons were then exported as ESRI Shapefiles, thereby making the outputs compatible with a wide range of GIS tools. The decision was made to segment non-drivable areas as opposed to *drivable* areas, because refuge islands could not be carved out of drivable polygons during the manual-annotation phase.

To increase image-segmentation accuracy in differentiating between drivable and non-drivable surfaces, two additional post-processing steps were applied. The first aimed to reduce drivable areas being incorrectly segmented as non-drivable. This was achieved by applying a buffer to all street centerlines encoded within OpenStreetMap. Specifically, any mask polygon with at least 10m$^2$ or 5% of its area overlapping with the centerline-buffer was removed, successfully eliminating many false positives (see **Figure 5**). Buffers from street centerlines were dynamically set given differences in width; roads tagged as 'primary' in OpenStreetMap had centerline buffers set at 2 meters, and untagged roads (primarily neighborhood streets) had centerline buffers set at 1.2 meters.

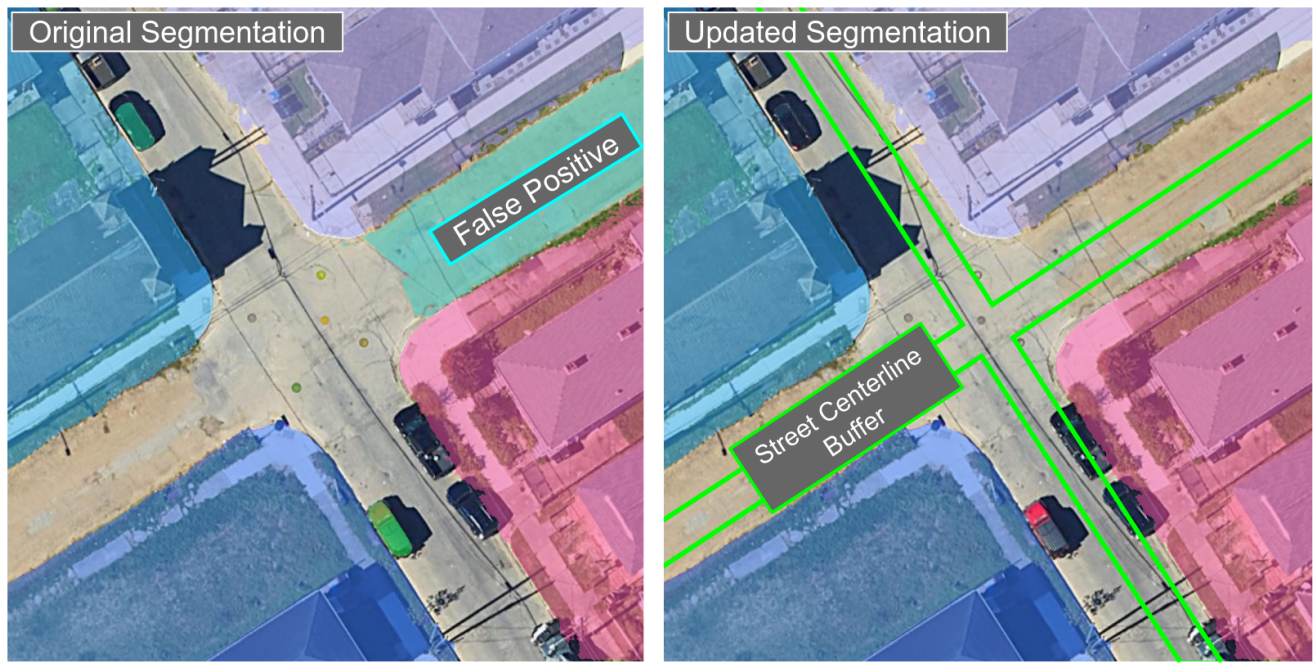

**Figure 5:** Intersection image with a false positive, non-drivable surface polygon (in turquoise, left). Segmentation improvement following the addition of a buffer (green lines, right) from the street centerline.

The second post-processing step aimed to include improperly omitted non-drivable areas (false negatives). To accomplish this, for each intersection image, all building footprints were extracted from OpenStreetMap as polygons and merged, creating a new non-drivable polygon that served as a substitute non-drivable mask, if one was not generated during image segmentation (see **Figure 6**).

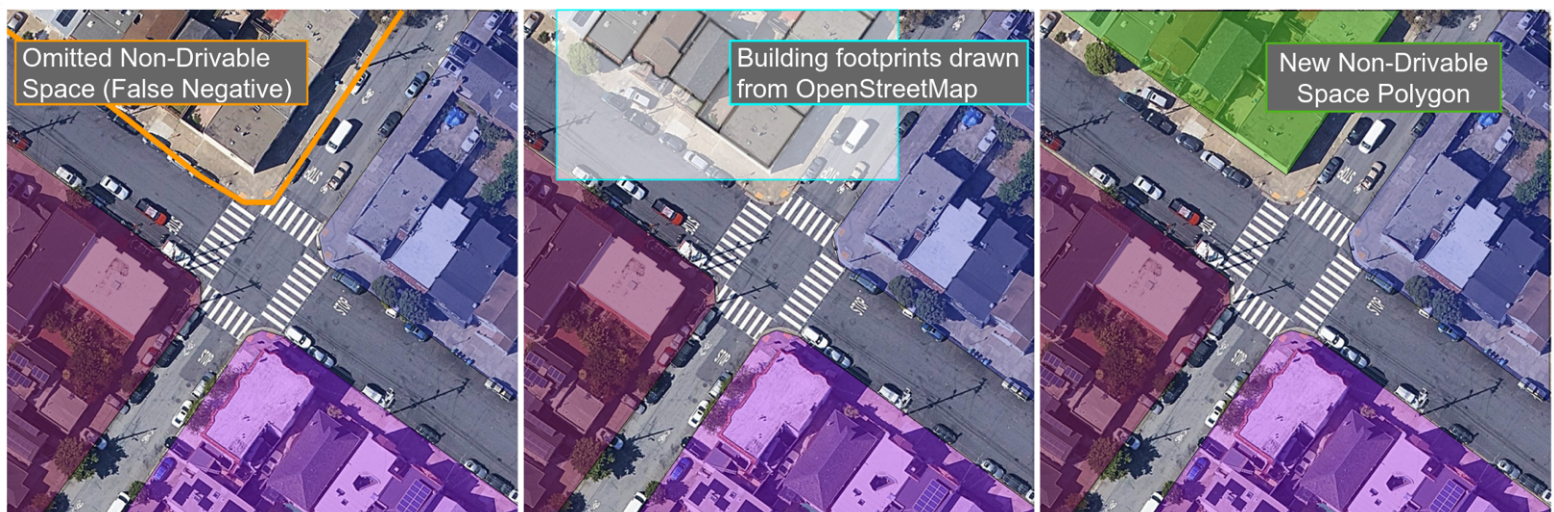

**Figure 6:** Example of an intersection image with an omitted non-drivable surface (false negative, left). The addition of merged building footprints drawn from OpenStreetMap (center). Newly-segmented mask polygon based (right).

## 2.2 Crossing Edge Extraction

OpenStreetMap contributors generally encode pedestrian crossings as edges (connected nodes). Given this convention, the Python package OSMnx was used to extract all pedestrian crossing edges for each city. However, some pedestrian crossings in OpenStreetMap are only encoded as single nodes, in the center of the street between sidewalks. In those cases, additional crossing edges were generated by matching each of those nodes to the closest edge in OpenStreetMap with the tag "highway." That allowed for construction of additional crossing edges; in some cities these newly-generated crossing edges comprised over 50% of the total (see *Supplementary Materials S2*). To avoid the issue of duplicating crossings where both a crossing edge *and* a crossing node were present, all edges with endpoints that were within 5m of each other's spatial location and within 5 degrees in terms of compass heading were merged into single edges. Though crossing edges in OpenStreetMap are not necessarily added by contributors in a way to precisely measure crossing distance (some do not reach sidewalks, whereas others end deep within sidewalks), they are of significant value as part of the grow-cut algorithm, described below.

## 2.3 Grow-Cut Algorithm

The final step of the analysis overlays the crossing edges extracted from OpenStreetMap atop the segmented intersection images. These edges provided a basis for applying a grow-cut algorithm, a common tool in image segmentation (Vezhnevets and Konouchine 2005). For each crossing edge, the grow-cut algorithm expanded the line by 5% from both endpoints, while maintaining its original bearing/orientation (see **Figure 7**). Next, the location at which the newly-grown edge and the non-drivable polygon intersected was identified. Finally, the algorithm cut each edge where it intersected with the non-drivable polygons, resulting in a final layer of edges that represent pedestrian crossings.

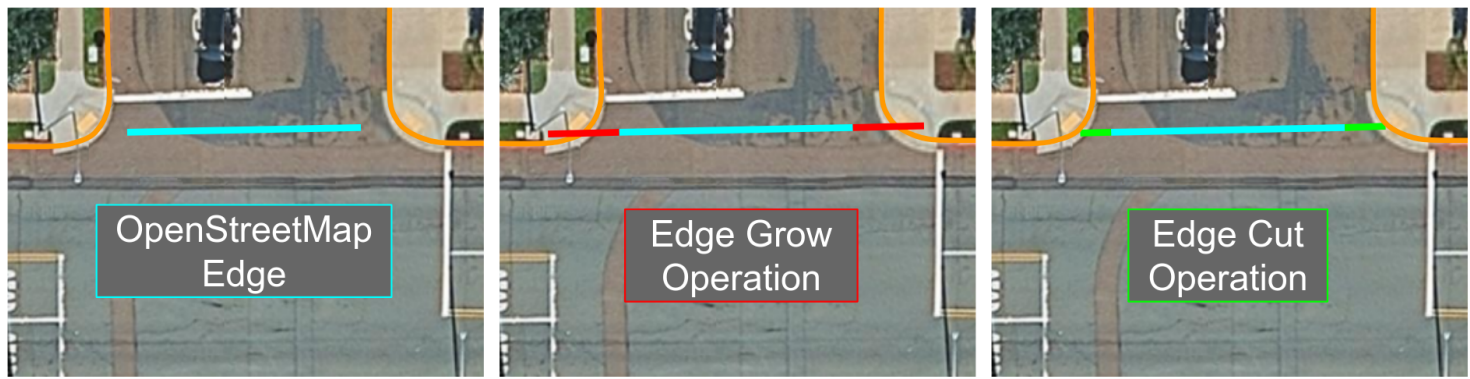

**Figure 7:** Grow-cut algorithm sequence, in which the crosswalk edge from OpenStreetMap (left) is first grown by a magnification factor (center), and then cut based on the boundary of the non-drivable space segmentation (right).

Two additional post-processing steps were applied after completion of the grow-cut algorithm. The first removed any remaining duplicate crossings by constructing a buffer around all edges by 6.5ft (2m). Then all buffered edges that intersected and were within 10 degrees compass heading of each other were merged. Second, crossing edges smaller than 6 ft (1.83m) and larger than 160 ft (61m) were removed. These bounds were set because visual inspection indicated there were effectively no pedestrian crossings shorter than six feet (the shortest being a crossing between a sidewalk and refuge island that is divided by a bike lane), and a very high proportion of crossings above 160 feet were determined to be erroneous as well. An expanded flow chart of the analysis is included in *Supplementary Materials S3*.

**3. Results**

The combination of intersection-image segmentation, crossing edge extraction from OpenStreetMap, and grow-cut algorithm application achieved automatic measurement of 808,377 pedestrian crossings throughout the 100 largest cities in the U.S. This approach computed crossing distances for each city in roughly one hour, with the bulk of the processing time being dedicated to image fetching and segmentation by SAM. Visual inspection of automated crossings demonstrates successful placement throughout a wide range of urban densities, land-uses, and intersection layouts (see **Figure 8)**.

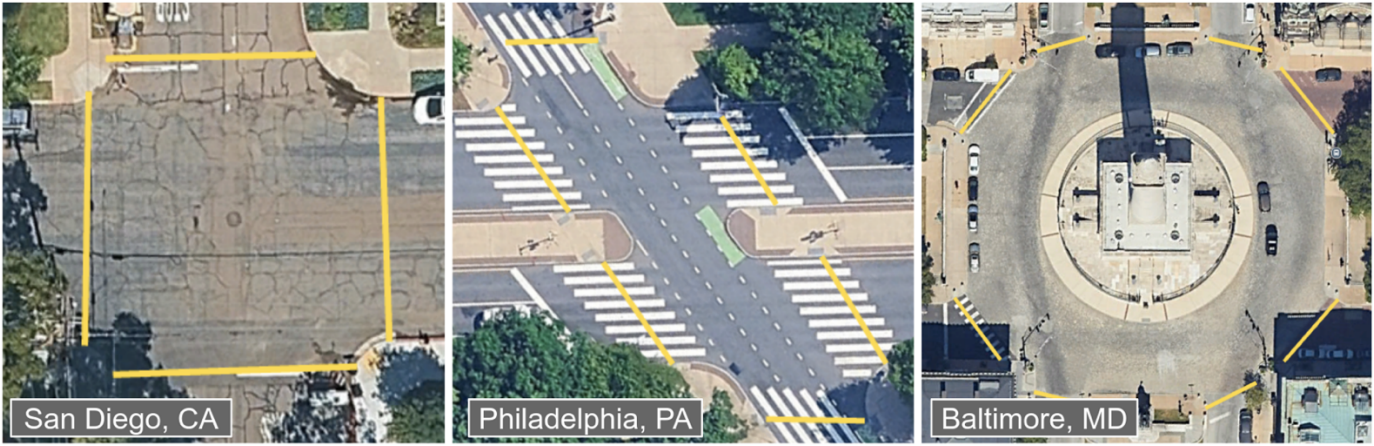

**Figure 8:** Automatically generated crossing distances at intersections of varying complexity in San Diego (left), Philadelphia (center), and Baltimore (right), including both marked and unmarked crossings.

### 3.1 Crossing Distance Accuracy

When compared to a citywide, manually-verified crossing-distance dataset of San Francisco (drawn from Moran and Laefer, 2024), automated crossing distances achieved an accuracy of 93%, with a median absolute error of 2 feet, 3 inches (0.69m). Comparing the distributions of crossing distance for both the manually-verified and automated datasets of San Francisco illustrates that accuracy remains high throughout the range (see **Figure 9**). In terms of coverage – the extent to which the automated process generated the *number and spatial extent* of crossings manually verified – the San Francisco comparison yielded 72% coverage, given availability of crossing edges in OpenStreetMap to draw on to seed the grow-cut algorithm.

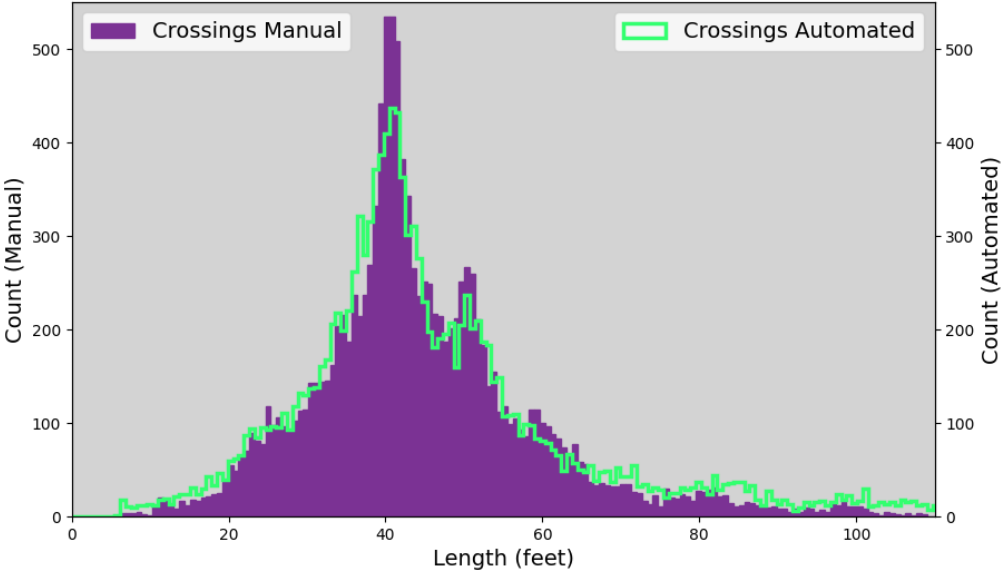

**Figure 9:** Distribution of pedestrian crossing distances in San Francisco, both via manual-verification (purple), and automated generation (green).

**3.2 Nationwide Analysis of Crossing Distance**

This approach enables the first nationwide assessment of pedestrian crossing distance, with a number of emergent patterns. First, viewed at the scale of an entire city, crossing distance exhibits two spatial signatures: clusters, which represent neighborhoods where crossing distance is similar throughout, and corridor effects, relating to specific arterials where crossing distance is nearly uniform along a linear stretch of road (see **Figure 10**). Indeed, in the maps of both Chicago and New York, clusters of shorter crossings are evident throughout, often separated from one another by wider arterial corridors (exhibiting longer crossings).

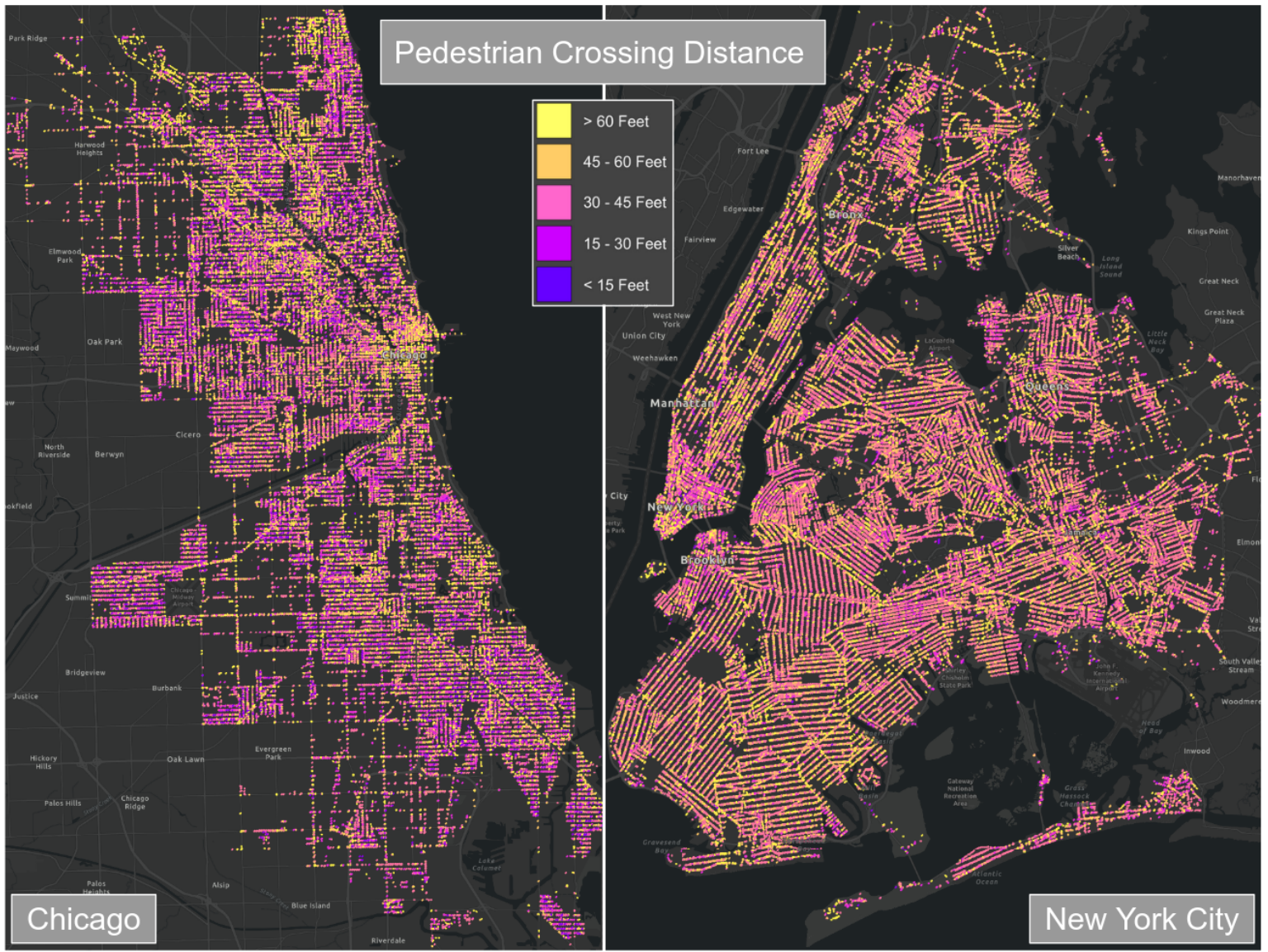

**Figure 10:** Maps of the spatial distribution of pedestrian crossing distances throughout Chicago (left), and New York City, excluding Staten Island (right).

Scaled to the 100 largest cities in the U.S., this also yields inter-city and regional comparisons; shorter median crossing distances clustered in the Northeast and Midwest (particularly in Michigan, Ohio, and Indiana), and longer median crossing distances are far more prevalent in the Southwest and West Coast (especially in the Dallas-Fort Worth area, southern Arizona, Southern Nevada, and Southern California; see **Figure 11**). To this point, Toledo, OH had the shortest median crossing distance (31 ft, 9.5 m) and North Las Vegas the longest (78 ft, 23.8 m).

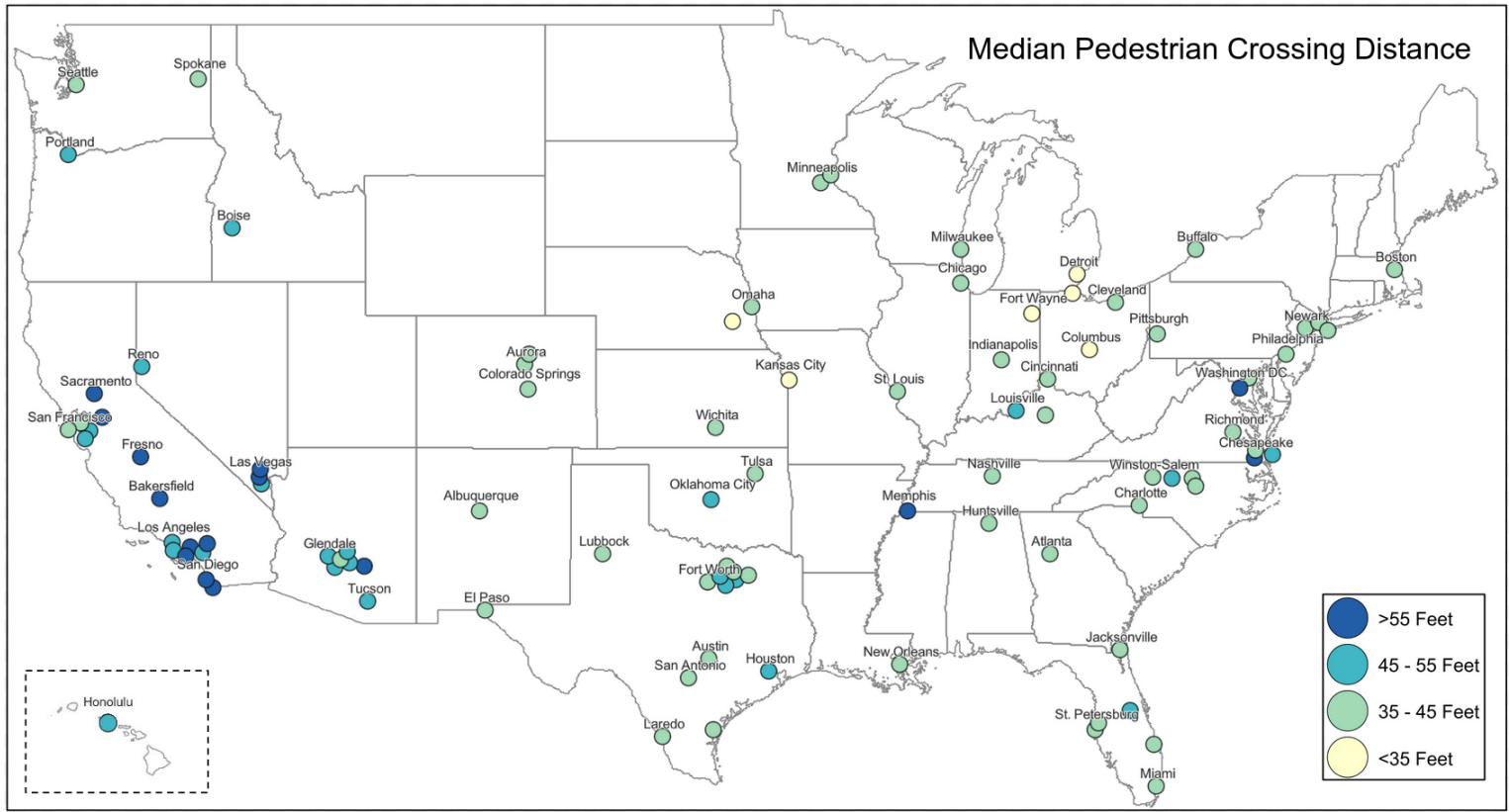

**Figure 11:** Map of median pedestrian crossing distance, by city.

By collapsing cities into three broad categories (1. Northeast and Midwest 2. South, Southeast, and Southwest, and 3. West Coast, Northwest, and Mountain West) it is further evident that crossings in Southern and Western cities, in the aggregate, are longer. They feature a peak of crossing distance around 40 feet (12.2m), whereas Northeast and Midwest cities have a peak far closer to 30 feet (9.1m, see **Figure 12a**). The sharp peak of the Northeast and Midwest cities value count for crossing distance also likely reflects those regions' streets primarily planned in rectilinear grids, relative to the more heterogeneous street layouts characteristic of more-sprawling municipalities (e.g. cul-de-sacs; Barrington-Leigh and Millard-Ball 2020), whose neighborhoods are often bracketed by wide multi-lane roads.

Comparing median crossing distance based on cities' year of incorporation highlights that older American cities tend to exhibit shorter median crossing distances, whereas younger cities tend to exhibit longer median crossing distances (see **Figure 12b**). Indeed, median crossing distance hovers around 40 feet (12.2m) for cities incorporated between 1600 and 1800, at which point there is a detectable and progressive trend of longer median crossing distances.

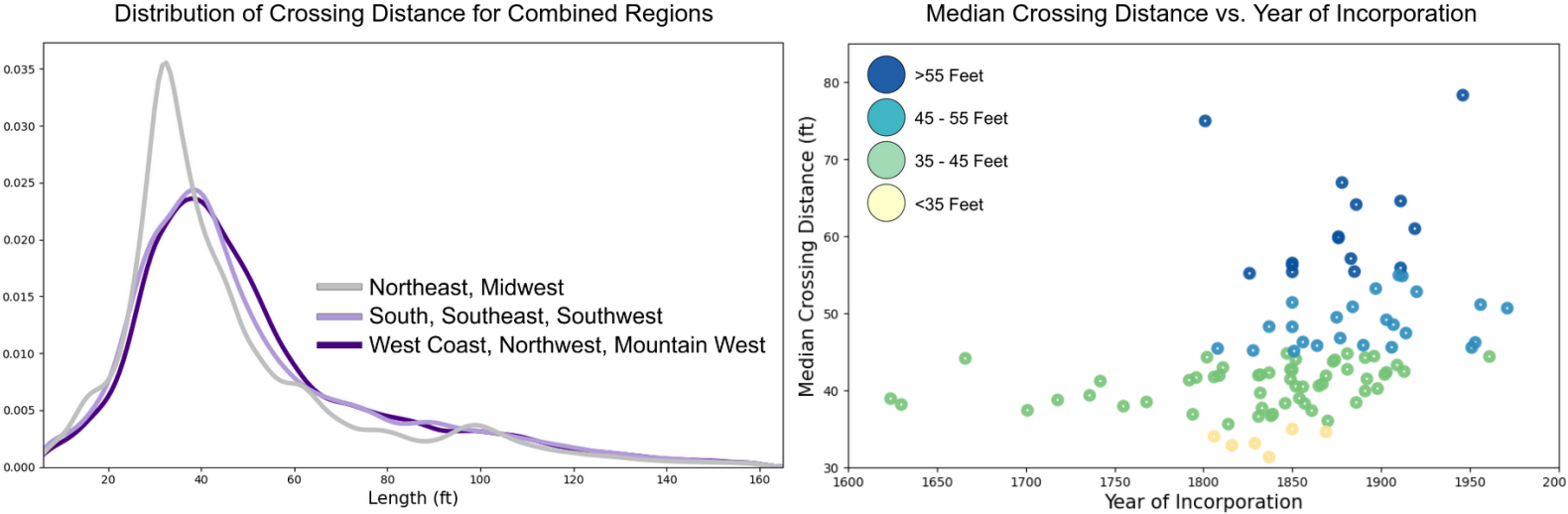

**Figure 12a:** KDE Plot of the distribution of crossing distance, for combined regions (left). **Figure 12b:** Scatterplot of median crossing distance vs. year of founding for the 100 largest U.S. cities (right).

## 4. Discussion and Conclusion

Deploying an advanced computer-vision model to segment satellite imagery of street intersections, in combination with information drawn from OpenStreetMap, generated crossing distance measurements throughout the 100 largest American cities (totaling approximately 800,000 crossings), with high accuracy when compared to manually-verified data. This establishes practitioner-relevant datasets using publicly-accessible data, and opens up analysis of this fundamental attribute of the pedestrian realm to new scales, both within and across cities. Unlike previous research which applied computer vision to pedestrian infrastructure, this approach captures both marked and *unmarked* crossings, a meaningful advancement given the latter comprises a large portion of total crossings in many American cities (in some cases at close to half of all intersections). In doing so, it demonstrates that crossing distance follows a universal pattern across U.S. cities: left-shifted normal distributions with a peak value count for a given distance that represents the average crossing along neighborhood streets, followed by declining values of increasing crossing distance. However, these datasets also exhibit differences associated with region and incorporation year. Specifically, median crossing distance is longer in cities in the South and West of the country, compared to those in the Northeast and Midwest. Chronologically, median crossing distance is also longer in cities incorporated after 1800, illustrating in a new way the increasing emphasis of car-centric streets and its effect on the pedestrian realm. In so doing, this study builds on scholarship exploring how street features maintain shared characteristics at multiple scales, and relate to the urban history of the United States (Jiang 2007; Barrington-Leigh and Millard-Ball 2015; Boeing 2021).

The implications of this study are threefold. First, in terms of practice, these datasets can provide planners, policymakers, and pedestrian-safety advocates with insights as to patterns in crossing distance in their own cities, potentially pinpointing neighborhoods, corridors, and specific intersections for traffic-calming investments. This creates the opportunity to combine these datasets with other locally-relevant attributes, such as speed limits, traffic signals, pedestrian volumes, historic crash data, and neighborhood demographics to further inform decision-making regarding street re-designs. In addition, there is now also the chance to determine if the associations between crossing distance and pedestrian safety, studied previously in smaller geographic contexts, hold at the national scale.

Second, there is likely strong interest in comprehensively measuring pedestrian crossing distance internationally; many nations have set ambitious goals of reducing pedestrian fatalities (Björnberg et al. 2022), and these methods can be adapted to cities globally, especially as high-resolution satellite imagery becomes available in more regions. Indeed, it would be fruitful to understand if such an approach could generate accurate measurements for cities in lower-income countries, including rapidly-urbanizing areas where road safety is a significant public health issue (Hyder et al. 2022).

Finally, this methodology represents the continued advancement of computer-vision models in the domain of urban research. These segmentation results portends a wide range of applications for identifying features within urban settings and at the sub-block scale. Future research could modify this methodology such that it captures bicycle infrastructure, transit-priority lanes, or other specific road markings. To that end, the full code underlying this analysis is available on GitHub for those who wish to proceed in these, or other directions.

There are several limitations to note. First, the use of crossing edges recorded in OpenStreetMap to seed the grow-cut algorithm inherently means that not all crossings were measured in each city. The comprehensiveness of pedestrian crossings in OpenStreetMap in American cities varies, a product of the volunteer-based nature of the map (Bennett 2010; Demetriou 2016). Though coverage was strong when compared to the manually-verified dataset for San Francisco (72% of all crossings present in the automated dataset), that figure likely varies across these 100 cases and is difficult to measure.

In terms of accuracy, there remain two areas for which future model improvements should be pursued. First, SAM has difficulty accurately segmenting intersection images where tree-cover is heavier, given canopies often visually block underlying sidewalks. For this reason, crossings in leafier neighborhoods are likely less accurate relative to areas with fewer trees, which can result in measurement errors. One potential method of reducing this source of error would be to segment satellite imagery of intersections drawn from *winter* months, when deciduous trees do not have leaves. Second, it is difficult to remove all false-positive crossing edges from these automated datasets, particularly in the case of highly-complex and idiosyncratic intersections, some of which include both crossing edges and separate crossing nodes in OpenStreetMap. Further modifying the grow-cut algorithm and post-processing steps will likely lead to improved outcomes on this front.

Lastly, Anchorage, Alaska, the 74th largest city in the United States by population was excluded from analysis because its very high latitude entails limited satellite imagery available from Google Maps. In its place, Frisco, Texas, the 101st largest city in the U.S. by population, was analyzed and incorporated into all subsequent analyses.

**Data Availability Statement:** Satellite imagery used in this study was drawn from Google Maps Tile Server (https://developers.google.com/maps/documentation/tile). The Segment Anything Model was built by Meta Inc. and is publicly available (https://segment-anything.com/). Code comprising this analysis can be accessed on GitHub: (https://github.com/agupta01/crossing-distances).

**Acknowledgements**
We thank the technology start-up company Modal (https://modal.com) for providing us with credits to conduct our analysis on their cloud-computing platform. In addition, we thank the NYU Center for Data Science for sponsoring the Capstone Projects, from which this study originated.

Supplementary Materials

**S1. List of 100 Largest U.S. Cities, by Population (2023 Estimate), and Assigned Region:**

1. New York, NY, 8,258,035, Northeast
2. Los Angeles, CA, 3,820,914, West Coast
3. Chicago, IL, 2,664,452, Midwest
4. Houston, TX, 2,314,157, South
5. Phoenix, AZ, 1,650,070, Southwest
6. Philadelphia, PA, 1,550,542, Northeast
7. San Antonio, TX, 1,495,295, South
8. San Diego, CA, 1,388,320, West Coast
9. Dallas, TX, 1,302,868, South
10. Jacksonville, FL, 985,843, Southeast
11. Austin, TX, 979,882, South
12. Fort Worth, TX, 978,468, South
13. San Jose, CA, 969,655, West Coast
14. Columbus, OH, 913,175, Midwest
15. Charlotte, NC, 911,311, Southeast
16. Indianapolis, IN, 879,293, Midwest
17. San Francisco, CA, 808,988, West Coast
18. Seattle, WA, 755,078, Northwest
19. Denver, CO, 716,577, Mountain West
20. Oklahoma City, OK, 702,767, South
21. Nashville, TN, 687,788, Southeast
22. Washington, DC, 678,972, Northeast
23. El Paso, TX, 678,958, South
24. Las Vegas, NV, 660,929, Mtn. West
25. Boston, MA, 653,833, Northeast
26. Detroit, MI, 633,218, Midwest
27. Portland, OR, 630,498, Northwest
28. Louisville, KY, 622,981, Midwest
29. Memphis, TN, 618,639, Southeast
30. Baltimore, MD, 565,239, Northeast
31. Milwaukee, WI, 561,385, Midwest
32. Albuquerque, NM, 560,274, Southwest
33. Tucson, AZ, 547,239, Southwest
34. Fresno, CA, 545,716, West Coast
35. Sacramento, CA, 526,384, West Coast
36. Mesa, AZ, 511,648, Southwest
37. Atlanta, GA, 510,823, Southeast
38. Kansas City, MO, 510,704, Midwest
39. Colorado Spr., CO, 488,664, Mtn. West
40. Omaha, NE, 483,335, Midwest
41. Raleigh, NC, 482,295, Southeast
42. Miami, FL, 455,924, Southeast
43. Virginia Beach, VA, 453,649, Southeast
44. Long Beach, CA, 449,468, West Coast
45. Oakland, CA, 436,504, West Coast
46. Minneapolis, MN, 425,115, Midwest
47. Bakersfield, CA, 413,381, West Coast
48. Tulsa, OK, 411,894, South
49. Tampa, FL, 403,364, Southeast
50. Arlington, TX, 398,431, South
51. Wichita, KS, 396,119, Midwest
52. Aurora, CO, 395,052, Mtn. West
53. New Orleans, LA, 364,136, South
54. Cleveland, OH, 362,656, Midwest
55. Honolulu, HI, 341,778, West Coast
56. Anaheim, CA, 340,512, West Coast
57. Henderson, NV, 337,305, Mtn. West
58. Orlando, FL, 320,742, Southeast
59. Lexington, KY, 320,154, South
60. Stockton, CA, 319,543, West Coast
61. Riverside, CA, 318,858, West Coast
62. Corpus Christi, TX, 316,595, South
63. Irvine, CA, 314,621, West Coast
64. Cincinnati, OH, 311,097, Midwest
65. Santa Ana, CA, 310,539, West Coast
66. Newark, NJ, 304,960, Northeast
67. Saint Paul, MN, 303,820, Midwest
68. Pittsburgh, PA, 303,255, Midwest
69. Greensboro, NC, 302,296, Southeast
70. Durham, NC, 296,186, Southeast
71. Lincoln, NE, 294,757, Midwest
72. Jersey City, NJ, 291,657, Southeast
73. Plano, TX, 290,190, South
74. Anchorage, AK, 286,075*
75. N. Las Vegas, NV, 284,771, Mtn. West
76. St. Louis, MO, 281,754, Midwest
77. Madison, WI, 280,305, Northeast
78. Chandler, AZ, 280,167, Southwest
79. Gilbert, AZ, 275,411, Southwest
80. Reno, NV, 274,915, Mountain West
81. Buffalo, NY, 274,678, Northeast
82. Chula Vista, CA, 274,333, West Coast

83. Fort Wayne, IN, 269,994, Midwest
84. Lubbock, TX, 266,878, South
85. Toledo, OH, 265,304, Midwest
86. St. Petersburg, FL, 263,553, Southeast
87. Laredo, TX, 257,602, South
88. Irving, TX, 254,373, South
89. Chesapeake, VA, 253,886, Southeast
90. Glendale, AZ, 253,855, Southwest
91. Winston-Salem, NC, 252,975, Southeast
92. Port St. Lucie, FL, 245,021, Southeast
93. Scottsdale, AZ, 244,394, Southwest
95. Boise, ID, 235,421, Mountain West
96. Norfolk, VA, 230,930, Southeast
97. Spokane, WA, 229,447, Northwest
98. Richmond, VA, 229,247, Southeast
94. Garland, TX, 243,470, South
99. Fremont, CA, 226,208, West Coast
100. Huntsville, AL, 225,564, Southeast
101. Frisco, TX, 225,007, South*

*Because of Anchorage, Alaska's high latitude, satellite imagery from Google Maps tile service is not available to the same extent as cities in the contiguous United States. For that reason, the 101st largest U.S. city by population, Frisco, Texas (population: 225,007, South), was used in its place.

**S2. Crossing Edges Extracted from OpenStreetMap, Original vs. Generated:**

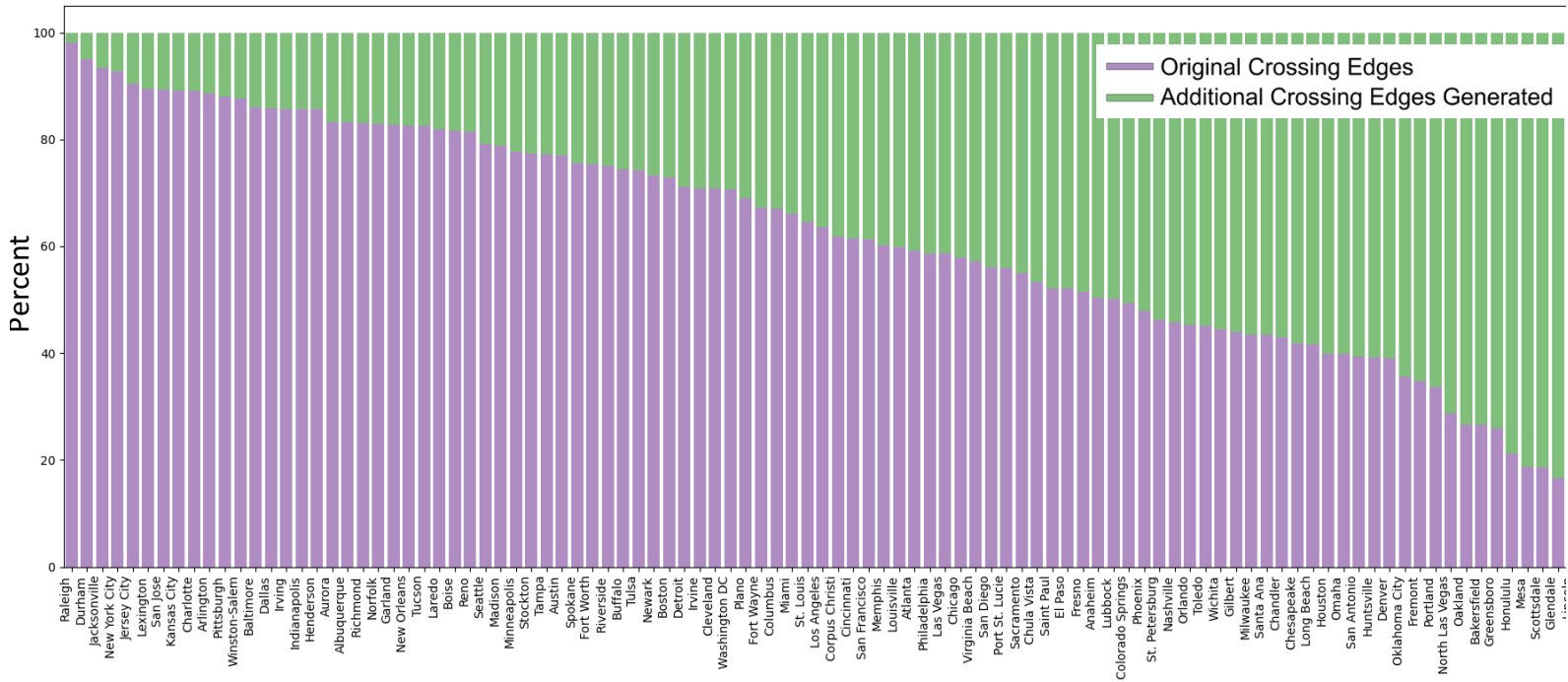

Because Frisco, TX, was incorporated into the data processing following the implementation of generating additional crossing edges from OpenStreetMap nodes, there is no 'original' figure for which to compare it to. Thus, it is omitted from this chart.

## S3. Flow-Chart for Data Processing

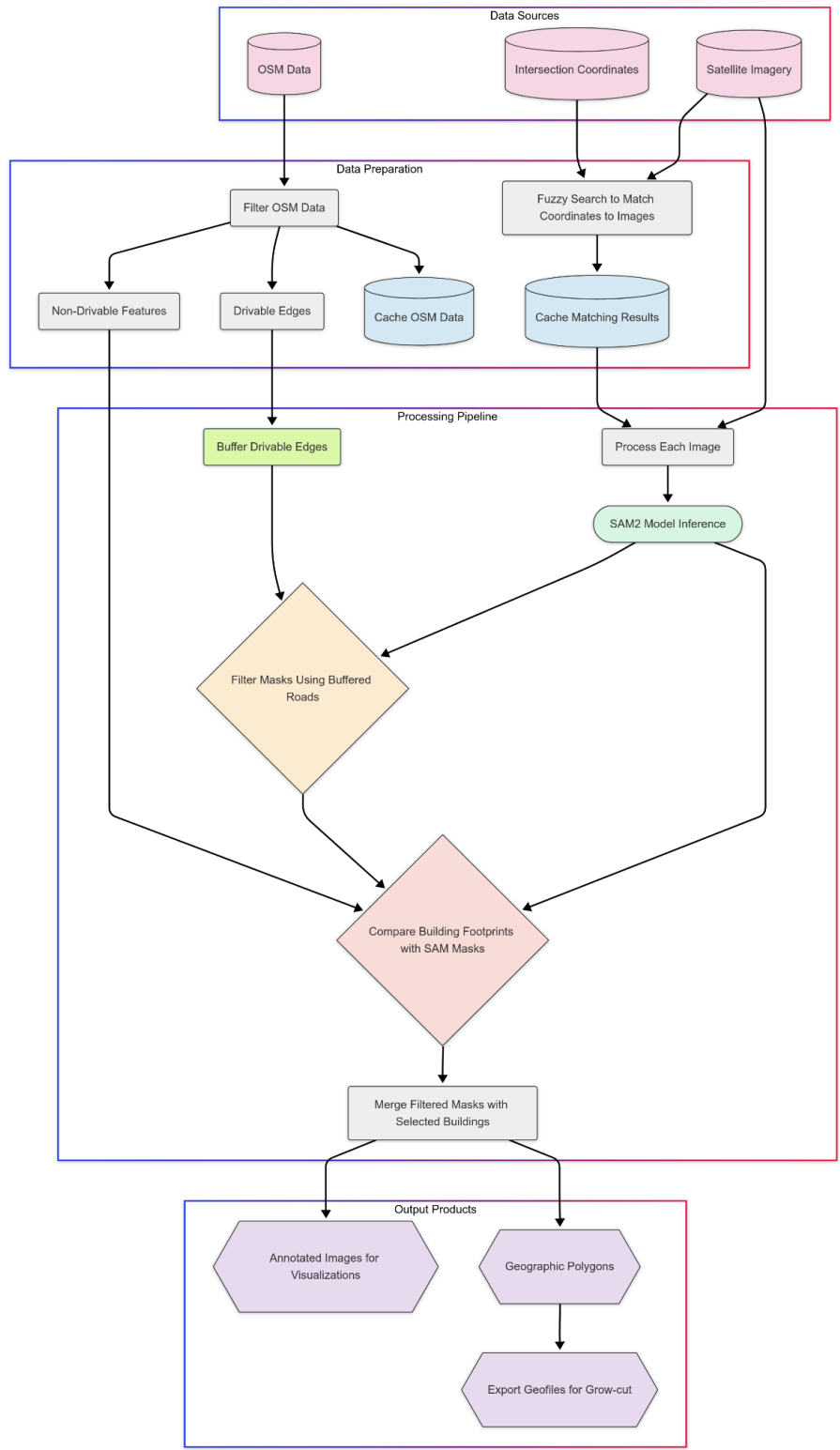